\documentclass[12pt]{article}
\usepackage[top=1in,bottom=1in,left=1.5in,right=1.5in]{geometry}
\usepackage{amsmath,amssymb}
\usepackage{latexsym}
\usepackage[dvips]{graphicx}

\newcommand{\C}{\mathbb{C}}
\newcommand{\N}{\mathbb{N}}
\newcommand{\R}{\mathbb{R}}
\newcommand{\Z}{\mathbb{Z}}
\renewcommand{\i}{\mathbf{i}}

\renewcommand{\a}{\mathbf{a}}

\newcommand{\x}{\mathbf{x}}
\newcommand{\X}{\mathbf{X}}
\newcommand{\y}{\mathbf{y}}

\newcommand{\bxi}{\mbox{\boldmath{$\xi$}}}
\newcommand{\0}{\mathbf{0}}

\newcommand{\cL}{\mathcal{L}}
\newcommand{\cM}{\mathcal{M}}

\newcommand{\rA}{\rm {A}}
\newcommand{\rE}{\rm {E\;}}

\renewcommand{\Pr}{\rm{Pr}}
\newcommand{\rS}{\rm{S}}

\newcommand{\lan}{\langle}
\newcommand{\ran}{\rangle}

\newcommand{\an}[1]{\lan#1\ran}
\newcommand{\hs}{\hspace*{\parindent}}
\newcommand{\proof}{\hs \textbf{Proof.\ }}
\newcommand{\tr}{\mathop{\mathrm{tr}}\nolimits}

\newcommand{\trans}{^\top}
\newcommand{\qed}{\hspace*{\fill} $\Box$\\}
\newcommand{\per}{\mathop{\mathrm{perm}}\nolimits}
\newcommand{\dete}{\mathop{\mathrm{det}_{\epsilon}}\nolimits}

\newcommand{\pfaf}{\mathop{\mathrm{pfaf}}\nolimits}
\newcommand{\Pf}{\mathop{\mathrm{Pf}}\nolimits}

\newtheorem{theo}{\bfseries \hs Theorem}[section]

\newtheorem{lemma}[theo]{\bfseries \hs Lemma}
\newtheorem{corol}[theo]{\bfseries \hs Corollary}

\numberwithin{equation}{section}

\begin{document}

\title{FPRAS for computing a lower bound for\\weighted matching polynomial
 of graphs }

\author{
  Shmuel Friedland \\
  Department of Mathematics, Statistics, and Computer Science,\\
  University of Illinois at Chicago\\
  Chicago, Illinois 60607-7045, USA}

 \date{April 12, 2007}

  \maketitle

 \begin{abstract}
 We give a fully polynomial randomized approximation scheme to
 compute a lower bound for the matching polynomial of any weighted graph
 at a positive argument.  For the matching polynomial of complete
 bipartite graphs with bounded weights these lower bounds are
 asymptotically optimal.
     \\[\baselineskip] 2000 Mathematics Subject
     Classification: 05A15, 05C70, 15A52, 68Q10.

 \end{abstract}


\section{Introduction}

 Let $G=(V,E)$ be an undirected graph, (with no self-loops), on the set of
 vertices $V$ and the set of edges $E$.
 A set of edges $M\subseteq E$ is called
 a \emph{matching} if no two distinct edges $e_1,e_2 \in M$ have a
 common vertex.  $M$ is called a $k$-\emph{matching} if $\#M=k$.
 For $k\in\N$ let $\cM_k(G)$ be the set of $k$-matchings in $G$.
 ($\cM_k(G)=\emptyset$ for $k>\lfloor \frac{\#V}{2}\rfloor$.)
 If $\#V=2n$ is even then an $n$-matching is called
 a \emph{perfect matching}.

 Let $\omega:E\to (0,\infty)$ be a weight function, which associate
 with edge $e\in E$ a positive weight $\omega(e)$.  We call $G_{\omega}=(V,E,\omega)$
 a weighted graph. Denote by $\iota$ the weight $\iota: E\to \{1\}$.
 Then $G$ can be identified with $G_{\omega}$.

 Let $M\in \cM_k(G)$.  Then the weight of the matching is defined as
 $\omega(M):=\prod_{e\in M} \omega(e)$.    \emph{The total weighted}
 $k$-\emph{matching} of $G_{\omega}$ is defined:
 $$\phi(k,G_{\omega}):=\sum_{M\in\cM_k(G)} \omega(M), k\in \N$$
 where $\phi(k,G_{\omega})=0$ if $\cM_k(G)=\emptyset$ for any $k\in
 \N$.  Furthermore we let $\phi(0,G_{\omega}):=1$.
 Note that $\phi(k,G_{\iota})=\#\cM_k(G)$, i.e. the
 number of $k$-matchings in $G$ for any $k\in\N$.
 The \emph{weighted matching polynomial} of $G_{\omega}$ is
 defined by:
 $$\Phi(t,G_{\omega}):=\sum_{k=0}^n \phi(k,G_{\omega})t^{n-k}, \quad n={\lfloor \frac{\#V}{2}\rfloor}.$$
 This polynomial is fundamental in
 the monomer-dimer model in statistical physics \cite{Bax3, HL},
 and for $\omega=1$ in combinatorics.  Note that if $\#V$ is even
 then $\Phi(0,G_{\omega})$ is the total weighted perfect matching of
 $G$.  (Some authors consider the polynomial $t^{\lfloor
 \frac{\#V}{2}\rfloor} \Phi(t^{-1},G_{\omega})$ instead of
 $\Phi(t,G_{\omega})$.)  It is known that nonzero roots of a weighted
 matching polynomial of $G$ are real and negative \cite{HL}.
 Observe that $\Phi(1,G_{\iota})$ the total number monomer-dimer coverings of $G$.

 Let $G$ be a bipartite graph, i.e., $V=V_1\cup V_2$ and
 $E\subset V_1\times V_2$.
 In the special case of a bipartite graph where $n=\#V_1=\#V_2$,
 it is well known that $\phi(n,G)$ is given as $\per B(G)$,
 the permanent of the incidence matrix $B(G)$ of the bipartite graph $G$.
 It was shown by Valiant that the computation of the
 permanent of a $(0,1)$ matrix is $\#$\textbf{P}-complete \cite{Val}.
 Hence, it is believed that the computation of the number of perfect
 matching in a general bipartite graph satisfying $\#V_1=\#V_2$ cannot
 be polynomial.

 In a recent paper Jerrum,
 Sinclair and Vigoda gave a \emph{fully-polynomial
 randomized approximation scheme} (\emph{fpras}) to compute the
 permanent of a nonnegative matrix \cite{JSV}.
 (See also Barvinok \cite{Bar}
 for computing the permanents within a simply exponential factor,
 and Friedland, Rider and Zeitouni \cite{FRZ} for
 concentration of permanent estimators
 for certain large positive matrices.)
 \cite{JSV} yields the existence a fpras to compute the total weighted
 perfect matching in a general bipartite graph satisfying
 $\#V_1=\#V_2$.  In a recent paper of Levy and the author
 it was shown that there exists fpras to
 compute the total weighted $k$-matchings for any bipartite graph $G$ and any
 integer $k\in[1,\frac{\#V}{2}]$.
 In particular, the generating matching polynomial of any bipartite
 graph $G$ has a fpras.  This observation can be used
 to find a fast computable approximation to the \emph{pressure} function, as discussed in
 \cite{FP}, for
 certain families of infinite graphs appearing in many models of statistical mechanics,
 like the integer lattice $\Z^d$ .

 The MCMC, (Monte Carlo Markov Chain), algorithm for computing the
 total weighted perfect matching in a general bipartite graph satisfying
 $\#V_1=\#V_2$, outlined in \cite{JSV}, can be applied to estimate
 the total weighted perfect matchings in a weighted non-bipartite graph
 with even number of vertices.  However the proof in
 \cite{JSV}, that shows this algorithm is frpas for bipartite
 graphs, fails for non-bipartite graphs.
 Similarly, the proof of
 concentration results given in \cite{FRZ} do not seem to work
 for non-bipartite graphs.  The technique introduced by Barvinok in
 \cite{Bar} to estimate the number of weighted perfect matching in
 bipartite graphs, does extend to the estimate of
 total weighted perfect matchings in a general non-bipartite graph
 with even number of vertices, when one uses real or complex Gaussian
 distribution.  (See the discussion in \S5.)

 In this paper we give a fpras
 for computing a lower bound $\tilde\Phi(t,G_{\omega})$ for
 the weighted generated function $\Phi(t,G_{\omega})$
 for a fixed $t>0$.  We show that this lower bound has a multiplicative error at
 most $\exp({N\min(\frac{a^2}{2t}}, C_1))$, see (\ref{presest}), where $a^2$ is the maximal weight of
 edges of $G$ and
 \begin{equation}\label{defC1}
 C_1=-\frac{1}{\sqrt{2\pi}}\int_{\R} \log (x^2)e^{-\frac{x^2}{2}} dx=1.270362845\ldots
 \end{equation}
 These estimates are similar in nature to heuristic computations of Baxter
 \cite{Bax1}, where he showed that his computation for the
 dimers on $\Z^2$ lattice are very precise
 away from only dimer configurations, i.e. perfect matchings.
 (The results of heuristic computations of Baxter were recently
 confirmed in \cite{FP}.)  We show that that for
 the matching polynomial of complete
 bipartite graphs with weights in $[b^2,a^2], 0<b\le a$, this lower bound
 is asymptotically optimal.

 We now describe briefly our technical results.
 With each weighted graph $G_{\omega}$ associate a skew symmetric
 matrix $A=[a_{ij}]_{i,j=1}^N \in \R^{N\times N},\;A\trans=-A$, where $N:=\#V$,
 as follows.
 Identify $E$ with $\an{N}:=\{1,\ldots,N\}$, and  each edge $e\in E$
 with the corresponding unordered pair $(i,j), i\ne j\in \an{N}$.
 Then $a_{ij}\ne 0$ if and only $(i,j)\in E$.  Furthermore for
 $1\le i< j\le N, (i,j)\in E$ $a_{ij}=\sqrt{\omega((i,j))}$.
 For $1\le i\le j\le \N$ let $x_{ij}$ be a set of $N\choose 2$
 independent random variables with
 \begin{equation}\label{normranvar}
 {\rE} x_{ij}=0, \quad {\rE}x_{ij}^2=1, \quad 1\le i \le j\le N.
 \end{equation}
 Let $\x:=(x_{11},\ldots,x_{1N},x_{22},\ldots,x_{NN})$.
 We view $\x$ as a random vector variable
 with values $\bxi=(\xi_{11},\ldots,\xi_{NN})\in \R^{N+1\choose 2}$.
 Let $Y_A$ be the following skew-symmetric random matrix
 \begin{equation}\label{defYA}
 Y_A:=[a_{ij}x_{\min(i,j)\max(i,j)}]_{i,j=1}^N\in \R^{N\times N}.
 \end{equation}

 A variation of the Godsil-Gutman estimator \cite{GG} states
 \begin{eqnarray}\label{GGE}
 {\rE}\det (\sqrt{t}I_N +Y_A)) =\Phi(t,G_{\omega}) \textrm{ if }
 N=\#V \textrm{ is even},\\
 \label{GGO}
 {\rE} \det (\sqrt{t}I_N +Y_A) =\sqrt{t}\Phi(t,G_{\omega}) \textrm{ if }
 N=\#V \textrm{ is odd}.
 \end{eqnarray}
 for any $t\ge 0$. Here $I_N$ stands for $N\times N$ identity
 matrix.

 We show the concentration of $\log\det (\sqrt{t}I_N +Y_A)$
 around
 \begin{equation}\label{deftilF}
 \log\tilde\Phi(t,G_{\omega}): ={\rE} \log\det (\sqrt{t}I_N +Y_A)
 \end{equation}
 using \cite{GZ}.  These concentration results show that
  $\tilde\Phi(t,G_{\omega})$
 has a fpras.  Jensen inequalities yield
 that $\tilde\Phi(t,G_{\omega})\le \Phi(t,G_{\omega})$.  Together
 with an upper estimate we have the following bounds:
 \begin{equation}\label{presest}
 \frac{1}{N}\log\tilde\Phi(t,G_{\omega})\le
 \frac{1}{N}\log\Phi(t,G_{\omega})\le
 \frac{1}{N}\log\tilde\Phi(t,G_{\omega})+ \min(\frac{a^2}{2t},C_1)
 \end{equation}
 where $a=\max |a_{ij}|$.  The above inequality hold also for $t=0$.
 (For $N$ even and $t=0$ this result is due to Barvinok
 \cite[\S7]{Bar}.)
 It is our hope that by refining the techniques we are using one can
 show that  $\Phi(t,G_{\omega})$ has fpras for any $t>0$.

 \section{Preliminary results}

 \begin{lemma}\label{ggfor}  Let $G=(V,E)$ be an undirected graph on
 $N$ vertices.  Let $\omega: V \to (0,\infty)$ be a given weight
 function.  Let $A=-A\trans\in \R^{n\times n}$ be the corresponding
 real skew symmetric matrix defined in \S1.  Assume that $x_{ij},
 i=1,\ldots,j, j=1,\ldots,N$ are $N+1\choose 2$ independent random
 variables, normalized by the conditions (\ref{normranvar}).
 Let $Y_A\in \R^{N\times N}$ be the skew symmetric real matrix
 defined by (\ref{defYA}).  Then (\ref{GGE}-\ref{GGO}) hold.
 \end{lemma}

 \proof  Let $\sqrt{t}=s$.  Observe first that $\det (sI_N +Y_A)$ is a sum of
 $N!$ monomials, where each monomial is of degree at most $2$ in the
 variables $x_{ij}$ for $i<j$ and of degree $m$ invariable $s$.  The
 total degree of each monomial is $N$.  The expected value of such a
 monomial is zero if at least the degree of one of the variables $x_{ij}$ is one.
 So it is left to consider the expected value of all monomials,
 where the degree if each $x_{ij}$ is $0$ or $2$, which are called
 nontrivial monomials.

 Assume first that $N$ is even.
 Observe that if a monomial contains $s$ of odd power than
 it must be linear at least in one $x_{ij}$.  Hence its
 expected value is zero.  Thus ${\rE}\det (sI_N +Y_A)$ is a polynomial
 in $s^2$.
 Consider a nontrivial monomial such that the power of $s$ is
 ${N-2m}$.  Note that this monomial is of the form $\tau s^{N-2m}
 \prod_{(i,j)\in M} \omega((i,j))x_{ij}^2$, for some $m$
 matching $M\in \cM_m$.  Here $(-1)^m\tau$ is the sign of the corresponding
 permutation $\sigma:\an{N}\to \an{N}$.
 Since $\sigma(i)=j, \sigma(j)=i$ for any edge $(i,j)\in M$,
 and $\sigma(i)=i$ for all vertices $i$ which are not covered by $M$ we
 deduce that $\tau=1$.  Hence the expected value of this monomial
 is  $s^{N-2m}\prod_{e\in M} \omega(e)$.  This proves (\ref{GGE}).
 The identity (\ref{GGO}) is shown similarly.  \qed

 Recall the following well known result:
 \begin{lemma}\label{skeweig}  Let $A=-A\trans\in\R^{N\times N}$ be
 a skew symmetric matrix.  Then $B:=\i A$, where $\i:=\sqrt{-1}$, is a
 hermitian matrix.  Arrange the eigenvalues of $B$ in a
 decreasing order:
 $\lambda_1(B)\ge\ldots\ge\lambda_{N}(B)$.
 Then
 \begin{equation}\label{skeweig1}
 \lambda_{N-i+1}(B)=-\lambda_i(B) \textrm{ for }
 i=1,\ldots,N.
 \end{equation}
 In particular
 \begin{equation}\label{skeweig2}
 \det(\sqrt{t}I_N+A)=\prod_{i=1}^N \sqrt{t+\lambda_i(B)^2}.
 \end{equation}
 \end{lemma}

 \proof  Clearly, $B$ is hermitian.  Hence all the eigenvalues of
 $B$ are real.  Arrange these eigenvalues in a decreasing order.
 So $-\i\lambda_j(B), j=1,\ldots,N$ are the eigenvalues of $A$.
 Since $A$ is real valued, the nonzero eigenvalues of $A$ must be
 in conjugate pairs.  Hence equality (\ref{skeweig1}) holds.
 Observe next that if $\lambda_k(A)=-\i \lambda_{k}(B)\ne 0$
 then
 $$(\sqrt{t} + \lambda_k(A))(\sqrt{t} + \lambda_{N-k+1}(A))=
 \sqrt{t+\lambda_k(B)^2} \sqrt{t+\lambda_{N-k+1}(B)^2}.$$
 As the eigenvalues of $\sqrt{t}I_N+A$ are $\sqrt{t}+\lambda_k(A),
 k=1,\ldots,N$ we deduce (\ref{skeweig2}).  \qed

 \section{Concentration for Gaussian entries}

 In this section we assume that each $x_{ij}$ is a normalized real
 Gaussian variable, i.e satisfying (\ref{normranvar}).  Recall that
 a function $f:\R\to \R$ is called \emph{Lipschitz} function, or \emph{Lipschtzian},
 if there exists $L\in
 [0,\infty)$ such that $\frac{|f(x)-f(y)|}{|x-y|}\le L$ for all
 $x\ne y \in\R$.  The smallest possible $L$ for a Lipschitz
 function is denoted by $|f|_{\cL}$.  Let $\rA_N\subset \R^{n\times
 n}, \i\rA_N\subset \C^{n\times n}$ denote the set of $N\times N$ real skew symmetric
 matrices,  and the set of $N\times N$ hermitian matrices of the
 form $\i A, A\in \rA_N$.  With each $A\in \rA_N$ we associate a
 weighted graph $G_{\omega}=(V,E,\omega)$, where $V=\an{N}, (i,j)\in
 V  \iff a_{ij}\ne 0, \omega((i,j))=|a_{ij}|^2$.  Denote by
 $a:=\max|a_{ij}|$.  To avoid the trivialities we assume that $a>0$.
 Note that $a^2$ is the maximal weight of the edges in $G_{\omega}$.
 Let $Y_A$ be the random skew symmetric matrix given by
  (\ref{defYA}) and denote by
 $X_A$ the random hermitian matrix
 $X_A:=\frac{1}{\sqrt{N}} \i Y_A$.

 Let $f:\R\to\R$ be a Lipschitz function.
 As in \cite{GZ} consider the following $F:\i \rA_N\to \R$ given by
 the trace formula:
 $$F(B)=\tr _N f(B):=\frac{1}{N} \sum_{i=1}^N f(\lambda_i(B)),
 \quad B\in \i\rA_N.$$
 Denote by ${\rE}\tr_N (f(X_A))$ the expected value of the function
 $\tr_N(f(X_A))$.  The concentration result \cite[Thm 1.1(b)]{GZ} states:

 \begin{equation}\label{GZR}
 {\Pr}(|\tr_N(f(X_A))-{\rE}\tr_N(f(X_A))|\ge r)\le
 2e^{-\frac{N^2r^2}{8a^2|f|_{\cL}^2}}
 \end{equation}
 (Recall that the normalized Gaussian distribution satisfies the log
 Sobolev inequality with $c=1$.)  We remark that since the entries
 of $X_A$ are either zero or pure imaginary one can replace the
 factor $8$ in the inequality (\label(\ref{GZR}) by the factor $2$.
 See for example the results in \cite[8.5]{Led}.

 \begin{lemma}\label{logconc}  Let $0\ne A=[a_{ij}]\in\rA_N, a=\max
 |a_{ij}|, t\in (0,\infty)$, $x_{ij}, 1\le i\le j\le N$ be
 independent Gaussian satisfying (\ref{normranvar}).  Let $Y_A\in \rA_N$ be
 the random skew symmetric matrix given by (\ref{defYA}).  Then
 \begin{equation}\label{logconc1}
 {\Pr}(|\log\det(\sqrt{t}I_N+Y_A)-{\rE}\log\det(\sqrt{t}I_N+Y_A)|\ge
 Nr)\le 2e^{-\frac{tNr^2}{2a^2}}.
 \end{equation}
 \end{lemma}
 \proof  Let $f_t(x):=\frac{1}{2} \log(\frac{t}{N}+x^2)$.  $f_t$ is
 differentiable and

 $$|(f_t)_{\cL}|=\max_{x\in\R} |f_t'(x)|=\frac{\sqrt{N}}{2\sqrt{t}}.$$
 Apply (\ref{GZR}) to $f_t$. Observe that the right-hand side
 of (\ref{GZR}) is equal to the right-hand side of (\ref{logconc1}).
 Use (\ref{skeweig2}) to deduce
 that
 \begin{eqnarray*}
 N\tr_N(f_t(X_A))= \sum_{i=1}^N \log \sqrt{\frac{t}{N} + \lambda_i(X_A)^2}=
 \sum_{i=1}^N\log \sqrt{\frac{t}{N} + \frac{|\lambda_i(Y_A)|^2}{N}}\\
 =-\frac{1}{2} N\log N +\log \prod_{i=1}^N \sqrt{t +
 |\lambda_i(Y_A)|^2}=
 -\frac{1}{2}N\log N + \log\det (\sqrt{t}I_N + Y_A).
 \end{eqnarray*}
 Hence the left-had sides of (\ref{GZR}) and
 (\ref{logconc1}) are equivalent.  \qed

 The following lemma is well known, e.g. \cite[p'1566]{FRZ}, and we bring its proof
 for completeness.

 \begin{lemma}\label{exest}  Let $U$ be a real random variable with
 a finite expected value ${\rE} U$.  Then $ e^{{\rE} U}\le
 {\rE} e^U$.
 Assume that the following condition hold
 \begin{equation}\label{probest}
 {\Pr} (U-{\rE} U\ge r)\le 2e^{_-Kr^2} \textrm{for each }
 r\in (0,\infty) \textrm{ and some } K>0.
 \end{equation}
 Then
 \begin{equation}\label{expin}
  e^{{\rE} U}\le {\rE} e^U\le e^{{\rE} U}
  (1+\frac{2e^{\frac{1}{4K}}}{\sqrt{K\pi}}).
  \end{equation}
 \end{lemma}

 \proof  Since $e^u$ is convex, the inequality $ e^{{\rE} U}\le {\rE} e^U$ follows
 from Jensen inequality.
 Let $\mu:={\rE} U$ and $F(u):={\Pr}(U\le u)$ be the
 cumulative distribution function of $U$.  We claim that
 \begin{equation}\label{eu2p1}
 {\rE} e^{U} \le e^{\mu} +\int_{\mu<u} e^u(1-F(u)) du.
 \end{equation}
 Clearly
 \begin{equation}\label{eu2p}
 {\rE} e^{U}=\int_{-\infty}^{\infty}  e^u dF(u) =
 \int_{u\le \mu} e^u dF(u) + \int_{\mu<u} e^u
 dF(u).
 \end{equation}
 Since $e^u\le e^{\mu}$ for $u\le \mu$ we deduce that
 $$\int_{u\le \mu} e^u dF(u)\le e^{\mu} F(\mu).$$
 We now estimate the second integral in the right-hand side of
 (\ref{eu2p}).  Recall that $F(u)$ is an
 nondecreasing function continuous from the right satisfying $F(+\infty)= 1$.
 Hence $e^u(F(u)-1)\le 0$ for all $u\in\R$. For any $R>\mu$
 use integration by parts to deduce
 \begin{eqnarray*}
 \int_{\mu<u \le R } e^u dF(u)=e^u(F(u) -1)|_{\mu}^{R}
 + \int_{\mu < u \le R} e^u(1-F(u))du
 \le \\ e^{\mu}(1-F(\mu))+\int_{\mu<u} e^s(1-F(u)) du.
 \end{eqnarray*}
 So
 $$ \int_{\mu<u  } e^u dF(u) \le  e^{\mu}(1-F(\mu))+\int_{\mu<u} e^u(1-F(u))
 du,$$
 and (\ref{eu2p1}) holds.

 Assume now that (\ref{probest}) holds. Thus
 $$1-F(u)={\Pr} (U>u)\le 2e^{-K(u-\mu)^2} \textrm{ for any }  u
 >\mu.$$

   Hence
 \begin{eqnarray*}
 \int_{\mu<u} e^u(1-F(u))du\le 2\int_{\mu<u} e^{u-K(u-\mu)^2}du\le\\
 2e^{\mu}\int_{-\infty}^{\infty} e^{-K(u-\mu-\frac{1}{2K})^2+\frac{1}{4K}}du=
 \frac{2e^{\mu}e^{\frac{1}{4K}}}{\sqrt{K\pi}}.
 \end{eqnarray*}
 Combine the above inequality with (\ref{eu2p1}) to deduce the right-hand side of
 (\ref{expin}).
 \qed

 \begin{corol}\label{exestc}  Let the assumptions of Lemma \ref{logconc}
 hold.  Then
 \begin{equation}\nonumber
 \frac{1}{N}\log\tilde\Phi(t,G_{\omega})\le
 \frac{1}{N}\log\Phi(t,G_{\omega})\le
 \frac{1}{N}\log\tilde\Phi(t,G_{\omega})
 +\frac{1}{N}\log(1+\frac{\sqrt{8N}a e^{\frac{a^2N}{2t}}}{\sqrt{\pi t}}).
 \end{equation}

 \end{corol}

 \section{FPRAS for computing $\log\tilde\Phi(t,G_{\omega})$}

 Let $B\in \R^{N\times N}$.  For $k\in\N$ denote by $\oplus_k B\in\R^{kN\times kN}$
 the block diagonal matrix ${\rm diag}(\underbrace{B,\ldots,B}_k)$.
 ($\oplus_k B$ is a direct sum of $k$ copies of $B$.)  Note that if
 $B\in \rA_N$ then $\oplus_k B\in \rA_{kN}$.
 Clearly,
 \begin{equation}\label{oplusid}
 \det (sI_{kN}+\oplus_k B) = (\det (sI_N +B))^k \textrm{ for any }
 B\in \R^{N\times N} \textrm{ and } s\in\R.
 \end{equation}
 Let $A\in \rA_N$, and $Y_A$ be the random matrix defined by
 (\ref{defYA}).
 By $Y_A(\bxi)$ we mean the skew symmetric matrix
 $[a_{ij}\xi_{\min(i,j)\max(i,j)}]_{i,j=1}^N$, which is a
 \emph{sampling} of $Y_A$.
 Let $x_{ij}, 1\le i \le j\le kN$ be $kN+1
 \choose 2$ normal Gaussian independent random variables.
 Consider the random matrix $Y_{\oplus_k A}$.  Then a sampling
 $$Y_{\oplus_k A}(\bxi),
 \bxi\in \R^{kN+1 \choose 2}={\rm diag} (Y_A(\bxi_1),\ldots,Y_A(\bxi_k)), \bxi_i\in \R^{N+1
 \choose 2}, i=1,\ldots,k$$
 is equivalent to $k$ sampling of $Y_A$.

 \begin{theo}\label{conopls}  Let $0\ne A=[a_{ij}]\in\rA_N, a=\max
 |a_{ij}|, t\in (0,\infty)$, $x_{ij}, 1\le i\le j\le N$ be
 independent Gaussian satisfying (\ref{normranvar}).  Let $Y_A\in \rA_N$ be
 the random skew symmetric matrix given by (\ref{defYA}).
 Let $Y_A(\bxi_1),\ldots,Y_A(\bxi_k)$ be $k$ samplings of $Y_A$.
 Then
 \begin{equation}\label{logconck}
 {\Pr}(|\frac{1}{k}\sum_{i=1}^k\log\det(\sqrt{t}I_N+Y_A(\bxi_i))-\log \tilde\Phi(t,G_{\omega})|\ge
 Nr)\le 2e^{-\frac{tkNr^2}{2a^2}}.
 \end{equation}
 In particular the inequality
 \begin{equation}\label{upestgap1}
 \frac{1}{N}\log\tilde\Phi(t,G_{\omega})\le
 \frac{1}{N}\log\Phi(t,G_{\omega})\le
 \frac{1}{N}\log\tilde\Phi(t,G_{\omega})
 +\frac{a^2}{2t}
 \end{equation}
 holds.

 Hence an approximation of $\tilde\Phi(t,G_{\omega})$ by
 $(\prod_{i=1}^k\det(\sqrt{t}I_N+Y_A(\bxi_i)))^{\frac{1}{k}}$ is a
 fully-polynomial randomized approximation scheme.
 \end{theo}

 \proof
 Use (\ref{oplusid}) to obtain
 $$\log\det (\sqrt{t}I_{kN} + Y_{\oplus_k A}(\bxi))=\sum_{i=1}^k \log\det (\sqrt{t}I_N+Y_A(\bxi_i))$$
 Hence
 \begin{equation}\label{ekform}
 {\rE}\log\det (\sqrt{t}I_{kN}+Y_{\oplus_k A}) = k  {\rE}\log\det ((\sqrt{t}I_N+ Y_{ A})=k \log
 \tilde\Phi(t,G_{\omega})
 \end{equation}
 Apply (\ref{logconc1}) to $Y_{\oplus_k A}$ to deduce (\ref{logconck}).
 Observe next that
 \begin{equation}\label{ekform1}
  {\rE}\det (\sqrt{t}I_{kN}+Y_{\oplus_k A}) =   {\rE}\det ((\sqrt{t}I_N+ Y_{
  A})^k=
 \Phi(t,G_{\omega})^k.
 \end{equation}
 Use Lemma \ref{exest} for the random variable $\log\det (\sqrt{t}I_{kN}+Y_{\oplus_k
 A})$ to deduce
 \begin{eqnarray*}
 \frac{1}{N}\log\tilde\Phi(t,G_{\omega})\le
 \frac{1}{N}\log\Phi(t,G_{\omega})\le
 \frac{1}{N}\log\tilde\Phi(t,G_{\omega})+\\
 +\frac{1}{kN}\log(1+\frac{\sqrt{8kN}a e^{\frac{a^2 kN}{2t}}}{\sqrt{\pi
 t}}).
 \end{eqnarray*}
 Let $k\to\infty$ to deduce (\ref{upestgap1}).

 We now show that (\ref{logconck}) gives fpras for computing
 $\tilde\Phi(t,G_{\omega})$ in sense of \cite{KL}.  Let
 $\epsilon,\delta
 \in (0,1)$.  Choose
 $$r=\frac{\epsilon}{2N}, \quad k=\lceil \frac{8a^2N \log \frac{4}{\delta}}{t\epsilon^2}
 \rceil .$$
 Then
 $${\Pr}(1-\epsilon < \frac
 {(\prod_{i=1}^k\det(\sqrt{t}I_N+Y_A(\bxi_i)))^{\frac{1}{k}}}
 {\tilde\Phi(t,G_{\omega})}<1+\epsilon)>1-\frac{\delta}{2}.$$
 Observe next that
 $${\Pr}(|x_{ij}|>\sqrt{2\log\frac{N^2k}{\delta}}\;)<\frac{\delta}{N^2k}.$$
 Hence with probability $1-\frac{\delta}{2}$ at least, the absolute of each off-diagonal of
 $Y_A(\bxi_i)), i=1,\ldots,k$ is bounded by
 $a\sqrt{2\log\frac{N^2k}{\delta}}$.  In this case all the entries
 of $\sqrt{t}I_N+Y_A(\bxi_i))$ are polynomial in $a,\sqrt{t},N,\frac{1}{\epsilon},\log
 \frac{1}{\delta}$.  The length of the storage of each entry is logarithmic in
 the above quantities.

 Finally observe that we need $O(N^3)$ to compute
 $\det(\sqrt{t}I_N+Y_A(\bxi_i))$.
 Hence the total number of computations for our estimate is of order
 $$t^{-1}a^2N^4 \epsilon^{-2}\log \delta^{-1}.$$  \qed

 The quantity $\frac{1}{N}\log\Phi(t,G_{\omega})$ can be viewed as
 the \emph{exponential growth} of $\log\Phi(t,G_{\omega})$ in terms
 of the number of vertices $N$ of $G$.
 Note that since the total number of matching of a graph $G$ is given by
 $\Phi(1,G_{\iota})$, Theorem \ref{conopls} combined with (\ref{presest}) yields that the
 exponential growth of the computable lower bound
 $\tilde\Phi(1,G_{\iota})$ differs by $\frac{1}{2}$ at most from
 the exponential growth of $\Phi(1,G_{\iota})$.
 Note that for complete graphs on $2n$, the exponential growth of the number of perfect
 matching matchings is of order $\log 2n -1$.  For $k$-regular
 bipartite graphs on $2n$ vertices the results of \cite{Fr, FP0} imply the
 inequality that for $n$ big enough the exponential growth of
 the total number of matchings is at least $\log k -1$.  Thus for
 graphs $G$ on $2n$ vertices containing,
 bipartite $k$-regular graphs on $2n$ vertices, with
 $k\ge 5$ and $n$ big enough,   $\tilde \Phi(1,G_{\iota})$ has a
 positive exponential growth.

 \section{Another estimate of $\log\Phi(t,G_{\omega})-\log\tilde\Phi(t,G_{\omega})$}

 \begin{lemma}\label{lgoexpin}  Let $X$ be a real Gaussian random
 variable.  Then
 \begin{equation}\label{lgoexpin1}
 \log {\rE} X^2 - {\rE}\log X^2\le C_1,
 \end{equation}
 where $C_1$ is given by (\ref{defC1}).
 Equality holds if and only if ${\rE}X=0$.

 \end{lemma}

 \proof  Clearly, it is enough to prove the lemma in the case
 $X=Y+a$, where $Y$ is a normalized by (\ref{normranvar}) and $a\ge
 0$.  In that case the left-hand side of (\ref{lgoexpin1}) is equal to

 $$g(a):=\log(1+a^2) -\frac{1}{\sqrt{2\pi}}\int_{\R} \log
 ((x+a)^2)e^{-\frac{x^2}{2}} dx.$$
 We used the software Maple to show that $f(a)$ is a decreasing
 function on $[0,\infty)$.  So $f(0)=C_1$ and $\lim_{a\to\infty}
 f(a)=0$.  This proves the inequality (\ref{lgoexpin1}).
 Equality holds if and only if $X=bY$ for some $b\ne 0$.  \qed

 Denote by $\rS_n\subset \R^{n\times n}$ the space of $n\times n$
 real symmetric matrices.
 A polynomial $P:\R^{n}\to \R$ is of degree $2$ if
 \begin{eqnarray*}
 P(\x)=\x\trans Q \x + 2\a\trans \x +b,\\
 \x=(x_1,\ldots,x_n)\trans, \a=(a_1,\ldots,a_n)\trans\in\R^n,
 Q\in\rS_n, b\in\R.
 \end{eqnarray*}
 (We allow here the case $Q=0$.)
 The quadratic form $P_h:\R^{n+1}\to \R$ induced by $P$ is given
 $$P_h(\y)=\y\trans Q_h\y, Q_h=\left[\begin{array}{cc} Q&\a\\
 \a\trans&b\end{array}\right]\in \rS_{n+1}, \y=(y_1,\ldots,y_{n+1})\trans.$$
 Clearly, $P(\x)=P_h((\x\trans,1)\trans)$.  $P$ is called a
 nonnegative polynomial if $P(\x)\ge 0$ for all $\x\in\R^n$.
 It is well known and a straightforward fact that $P$ is nonnegative
 if and only if $Q_h$ is a nonnegative definite matrix.

 The following lemma is a generalization of \cite[Thm 4.2, (1)]{Bar}.

 \begin{lemma}\label{bargen}  Let $P:\R^n\to\R$ be a nonzero
 nonnegative quadratic polynomial.  Let $X_1,\ldots,X_n$ be
 $n$-Gaussian random variables, and denote $\X:=(X_1,\ldots,X_n)\trans$.  Then
 \begin{equation}\label{bargen1}
 {\rE} \log P(\X)\le \log {\rE} P(\X)\le
 {\rE} \log P(\X) +C_1,
 \end{equation}
 where $C_1$ is given by (\ref{defC1}).
 \end{lemma}

 \proof We may assume without a loss of generality that ${\rE} P=1$.
 In view of the concavity of $\log$ we need to show the right-hand
 side of (\ref{bargen1}).
 Since $Q_h$ is nonnegative definite it follows that
 \begin{eqnarray*}
 P(\x)=\sum_{i=1}^m \lambda_i(\a_i\trans\x+b_i)^2,\;
 \a_i\in\R^n, b_i\in\R, \lambda_i>0, i=1,\ldots,m, \\
 {\rE} (\a_i\trans \X + b_i)^2=1,\; i=1,\ldots
 m,\quad
 \sum_{i= 1}^m \lambda_i=1.
 \end{eqnarray*}
 Note that one can have at most one $\a_i=\0$, and in that case then
 $b_i^2=1$.
 The concavity of $\log$ yields
 $$\log P(\X)\ge \sum _{i=1}^m \lambda_i \log (\a_i\trans \X + b_i)^2. $$
 (We assume that $\log 0=-\infty$.)
 Note that if $\a_i\ne 0$ then $\a_i\X+b_i$ is Gaussian.  Lemma \ref{lgoexpin1}
 yields ${\rE}\log P(\X)\ge -C_1$.  \qed

 \begin{theo}\label{barvthm}
 Let the assumptions of Theorem \ref{conopls} hold.  Then
 (\ref{presest}) holds.

 \end{theo}

 \proof In view of (\ref{upestgap1}) it is left to show
 \begin{equation}\label{presest1}
\log\Phi(t,G_{\omega})\le
 \log\tilde\Phi(t,G_{\omega})+ (N-1)C_1.
 \end{equation}
 Let $A=[a_{ij}]_{i,j=1}^{n}\in \rA_N$.  Recall that $\det A=(\pfaf
 A)^2$, where $\pfaf A$ is the pfaffian.  (So $\pfaf A=0$ if $n$ is
 odd.)
 Let $\a_i=(a_{1i},\ldots,a_{(i-1)i})\trans\in
 \R^{i-1}, i=2,\ldots,n$.  We view $\pfaf A$ as multilinear polynomial
 $\Pf(\a_2,\ldots,\a_n)$ of total degree $\frac{n}{2}$, which is linear in each vector variable
 $\a_i$.  (Any polynomial of noninteger total degree is zero
 polynomial by definition.)

 Denote by $Q_{k,n}$ the set of subsets of $\an{n}$ of
 cardinality $k\in [1,n]$.  Each $\alpha\in Q_{k,n}$ is viewed as
 $\alpha=\{i_1,\ldots,i_k\}, 1\le i_1<\ldots<i_k\le m$.
 For any matrix $B=[b_{ij}]\in\R^{n\times n}$ and $\alpha\in Q_{k,n}$ we define
 $B[\alpha|\alpha]\in \R^{k\times k}$ as the principal submatrix
 $[b_{\alpha_i\alpha_j}]_{i,j=1}^k$.  Then for $A=[a_{ij}]\in \rA_n$
 denote
 $$\Pf_{\alpha}(\a_2,\ldots,\a_n):=\pfaf A[\alpha|\alpha].$$
 Then $\Pf_{\alpha}(\a_2,\ldots,\a_n)$ is a multilinear polynomial
 of total degree $\frac{k}{2}$, which is linear in each $\a_i$.
 Hence
 \begin{equation}\label{detid}
 \det(sI_N+A)=s^N+\sum_{k=1}^n s^{N-k}\sum_{\alpha\in Q_{k,n}}
 \Pf_{\alpha}(\a_2,\ldots,\a_N)^2, \textrm{ for any } A\in \rA_N.
 \end{equation}
 View $\a_i\in\R^{i-1}$ as a variable while all other $\a_2,\ldots,\a_N$ are fixed.
 Then for $s\ge 0$ the above polynomial is quadratic and nonnegative.
 Group the $N\choose 2$ independent normalized random Gaussian
 variables $X_{ij}, 1\le i<j\le N$ into $N-1$ random vectors
 $\X_i:=(X_{1i},\ldots,X_{(i-1)i})\trans, i=2,\ldots,N$.
 Consider now $Y_A$.  Let
 $$P(\X_2,\ldots,\X_N):=\det (\sqrt{t}I_N +Y_A) \quad t\ge 0.$$
 Then $P(\X_2,\ldots,\X_N)$ is a nonnegative quadratic polynomial in
 each $\X_j, j=2,\ldots,N$.  Denote by ${\rE}_i$ the expectation with respect to
 the variables $X_{1i},\ldots, X_{(i-1)i}$.  (\ref{detid}) yields
 that
 $$P_{i}(\X_2,\ldots,\X_i):={\rE}_{i+1}\ldots{\rE_{N}}
 P(\X_2,\ldots,\X_N)$$
 is a nonnegative quadratic polynomial in
 each $\X_j, j=2,\ldots,i$.  Lemma \ref{bargen} yields
 $$\log{\rE}_i P_i(\X_2,\ldots,\X_i)\le {\rE_i} \log
 P_i(\X_2,\ldots,\X_{i})+C_1, \quad i=2,\ldots,N.$$
 Hence
 \begin{eqnarray*}
 \log\Phi(t,G_{\omega})=\log {\rE}_2 P_2(\X_2)\le {\rE_2}\log
 P_2(\X_2)+C_1\le \\
 {\rE}_2{\rE_3}\log P_3(\X_2,\X_3) + 2C_1\le \ldots\le\\
 {\rE}_2{\rE_3}\ldots{\rE}_N\log P(\X_2,\X_3,\ldots,\X_N) +
 (N-1)C_1=\\
 \log\tilde\Phi(t,G_{\omega})+(N-1)C_1.
 \end{eqnarray*}
 \qed

 \section{Bipartite graphs}
 Assume that $G=(V,E)$ is a bipartite graph.  So $V=V_1\cup V_2,
 E\subset E_1\times E_2$ and $N=m+n$.
 Assume for convenience of notation that
 $m:\#V_1\le n:=\#V_2$.  Thus $E\subset \an{m}\times\an{n}$, so each $e\in E$ is
 identified uniquely with $(i,j)\in \an{m}\times \an{n}$.  Let
 $C=[c_{ij}]\in\R^{m\times n}$ be the weight matrix associated with
 the weights $\omega: E\to (0,\infty)$.  So $c_{ij}=0$ if
 $(i,j)\not\in E$ and $c_{ij}=\sqrt{\omega(i,j)}$ if $(i,j)\in E$.
 Let $x_{ij},i=1,\ldots,m,j=1,\ldots,n$ be $mn$ independent
 normalized real Gaussian variables.  Let $U_C=:[c_{ij}x_{ij}]\in \R^{m\times n}$ be
 a random matrix.  Then the skew symmetric matrix $A$ associated with $G_{\omega}$ is
 given by and the corresponding random matrices $Y_A,X_A$ are given
 as
 \begin{equation}\label{defAYX}
 A=\left[\begin{array}{cc} 0&C\\-C\trans&0\end{array}\right],
 Y_A=\left[\begin{array}{cc} 0&U_C\\-U_C\trans&0\end{array}\right],
 X_A=\frac{\i}{\sqrt{m+n}}Y_A .
 \end{equation}
 Denote by
 \begin{equation}\label{svuc}
 \sigma_1(U_C)\ge\ldots\ge\sigma_m(U_C)\ge 0
 \end{equation}
 be the first $m$ singular values of $U_C$.
 Then the eigenvalues of $Y_A$ consists of $n-m$ zero eigenvalues
 and the following $2m$ eigenvalues:
 \begin{equation}\label{2meigy}
 \pm \i\sigma_1(U_C),\ldots,\pm\i\sigma_m(U_C).
 \end{equation}
 Hence
 \begin{equation}\label{detyabp}
 \det (\sqrt{t}I_{m+n} + Y_A)=t^{\frac{n-m}{2}}\prod_{i=1}^m
 (t+\sigma_i(U_C)^2).
 \end{equation}
 In \cite{FRZ} the authors considered the random matrix
 $V_C:=U_C U_C\trans \in \R^{m\times m}$.
 Note that the eigenvalues of $V_C$ are
 \begin{equation}\label{eigz}
 \sigma_1^2(U_C)\ge\ldots \ge \sigma_m^2(U_C).
 \end{equation}
 Furthermore, one has the equality
 ${\rE} \det V_C=\phi(m,G_{\omega})$.  Let $K_{m,n}$ be the complete
 bipartite graph on $V_1=\an{m}, V_2=\an{n}$ vertices.  Assume that
 $ 1\le m\le n$.  Let $0<b\le a$ be fixed.  Denote by
 $\Omega_{m,n,[b^2,a^2]}$ the sets of  all weights
 $\omega:\an{m}\times\an{n}\to [b^2,a^2]$.  Recall that each
 $\omega\in \Omega_{m,n,[b^2,a^2]}$ induces the positive matrix
 $C(\omega)=[c_{ij}(\omega)] \in \R^{m\times n}$, where
 $c_{ij}(\omega)\in [b,a]$.  It was shown in \cite{FRZ} that
 $\frac{1}{n}\log\det V_{C(\omega)}$  concentrates at
 $\frac{1}{n}\log\phi(m,K_{m,n,\omega})$ with probability $1$
 as $n\to\infty$.  More precisely
 \begin{equation}\label{frzcon}
 \limsup_{n\to\infty} \sup_{m\le n, \omega\in\Omega_{m,n,[b^2,a^2]}}
 {\Pr}(\frac{1}{n}|\log\det V_{C(\omega)}-
 \log\phi(m,K_{m,n,\omega})|>\delta)=0
 \end{equation}
 for any $\delta>0$.

 \begin{theo}\label{conoplsbp}  Let $0<b\le a$ be given.  For
 $\omega\in\Omega_{m,n,[b^2,a^2]}$ let $C(\omega)$ be a positive
 ${m\times n}$ matrix defined above and $A(\omega)\in \rA_{m+n}$ be
 given by (\ref{defAYX}), ($C=C(\omega)$).
 Assume that $x_{ij}, 1\le i\le j\le (m+n)$
 are independent Gaussian satisfying (\ref{normranvar}).  Let $Y_A\in \rA_N$ be
 the random skew symmetric matrix given by (\ref{defYA}).  Then for
 any $t>0$

 \begin{equation}\label{conoplsbp1}
 \limsup_{n\to\infty} \sup_{m\le n, \omega\in\Omega_{m,n,[b^2,a^2]}}
 {\Pr}(\frac{1}{m+n}|
 \log\det(\sqrt{t}I_N+Y_A)-\log\Phi(t,K_{m,n,\omega})|>\delta)=0
 \end{equation}
 Equivalently
 \begin{equation}\label{conoplsbp2}
 \limsup_{n\to\infty} \sup_{m\le n, \omega\in\Omega_{m,n,[b^2,a^2]}}
 \frac{1}{m+n}(\log\Phi(t,K_{m,n,\omega})
 -\log\tilde\Phi(t,K_{m,n,\omega}))=0.
 \end{equation}

 \end{theo}

 \proof  Our proof follows the arguments in \cite{FRZ}, and we point
 out the modifications that one has to make.  Let $N=m+n$.  Since
 $1\le m\le n$ we have that $\frac{1}{2n}\le \frac{1}{N}< \frac{1}{n}$.
 (\ref{logconck}) with $k=1$ implies:
 \begin{equation}\label{conoplsbp3}
 \limsup_{n\to\infty} \sup_{m\le n, \omega\in\Omega_{m,n,[b^2,a^2]}}
 {\Pr}(\frac{1}{m+n}|\log
 \det(\sqrt{t}I_N+Y_A)-\log\tilde\Phi(t,K_{m,n,\omega})|>\delta)=0
 \end{equation}
 Thus it is enough to show equality (\ref{conoplsbp2}).

 Denote by
 $X_A$ the random hermitian matrix
 $X_A:=\frac{1}{\sqrt{N}} \i Y_A$.
 For $\epsilon >0$ define
 \begin{eqnarray*}
 \dete(\sqrt{t}I_N+ Y_N):=\prod_{i=1}^N
 \sqrt{t+\max(|\lambda_i(Y_N)|,\sqrt{N}\epsilon)^2},\\
 \dete(\frac{\sqrt{t}}{\sqrt{N}}I_N-\i X_N):=\prod_{i=1}^N
 \sqrt{\frac{t}{N}+\max(|\lambda_i(X_N)|,\epsilon)^2}.
 \end{eqnarray*}
 Clearly,
 \begin{equation}\label{eqdete}
 \dete(\sqrt{t}I_N+ Y_N)=N^{\frac{N}{2}}
 \dete(\frac{\sqrt{t}}{\sqrt{N}}I_N-\i X_N).
 \end{equation}
 Let $f_{N,t,\epsilon}(x):=\frac{1}{2}\log
 (\frac{t}{N}+\max(|x|,\epsilon)^2)$.  Then
 $$|f_{N,t,\epsilon}|_{\cL}\le \frac{1}{\epsilon} \textrm{ for }
 N\ge \frac{t}{\epsilon^2}.$$
 In what follows we assume that $N\ge \frac{t}{\epsilon^2}$.
 Observe next that
 $$\frac{1}{N}\log  \dete(\frac{\sqrt{t}}{\sqrt{N}}I_N-\i X_N) =\tr _N
 f_{N,t,\epsilon}(X_A).$$
 Combine the concentration inequality (\ref{GZR}) with (\ref{eqdete})
 to obtain
 \begin{equation}\label{GZRe}
 {\Pr}(|\frac{1}{N}(\log  \dete(\sqrt{t}I_N+ Y_N)-
 {\rE}\log  \dete(\sqrt{t}I_N+ Y_N))|\ge r)\le
 2e^{-\frac{N^2r^2\epsilon^2}{8a^2}}
 \end{equation}
 Let
 \begin{equation}\label{chsen}
 \epsilon_N=\frac{1}{(\log N)^2}.
 \end{equation}
 Note that for a fixed $t$ one has $N\ge \frac{t}{\epsilon_N^2}$ for
 $N>>1$.  Hence
 $$\limsup_{N\to\infty}{\Pr}(\frac{1}{N}|\log  \textrm{det}_{\epsilon_N}(\sqrt{t}I_N+ Y_N)-
 {\rE}\log  \textrm{det}_{\epsilon_N}(\sqrt{t}I_N+ Y_N)|\ge \delta)=0$$
 for any $\delta>0$.
 As in \cite[Prf. of Lemma 2.1]{FRZ} use (\ref{GZRe}) and Lemma \ref{exest} to deduce that
 $$\lim_{N\to\infty}\frac{1}{N}(\log  {\rE}\textrm{det}_{\epsilon_N}(\sqrt{t}I_N+ Y_N)-
 {\rE}\log  \textrm{det}_{\epsilon_N}(\sqrt{t}I_N+ Y_N))=0,$$
 which is equivalent to
 \begin{equation}\label{GZRi0}
 \lim_{N\to\infty}\frac{1}{N}(\log{\rE}
 \textrm{det}_{\epsilon_N}(\frac{\sqrt{t}}{\sqrt{N}}I_N-\i
 X_N)-{\rE}\log
 \textrm{det}_{\epsilon_N}(\frac{\sqrt{t}}{\sqrt{N}}I_N-\i
 X_N))=0.
 \end{equation}
 It is left to show that under the assumption of the theorem
 \begin{equation}\label{GZRi1}
 \lim_{N\to\infty}\frac{1}{N}(\log  {\rE}\textrm{det}_{\epsilon_N}(\sqrt{t}I_N+
 Y_N)-\log  {\rE}\det(\sqrt{t}I_N+
 Y_N))=0.
 \end{equation}
 Clearly, the above claim is equivalent to
 \begin{equation}\label{GZRi2}
 \lim_{N\to\infty}\frac{1}{N}(\log
 {\rE}\textrm{det}_{\epsilon_N}(\frac{\sqrt{t}}{\sqrt{N}}I_N-\i
 X_N)-\log
 {\rE}\det(\frac{\sqrt{t}}{\sqrt{N}}I_N-\i
 X_N))=0.
 \end{equation}
 To prove the above equality we use the results of \cite{FRZ}.
 First observe that $X_N$ has at least $n-m$ eigenvalues which are
 equal to zero, while the other $2m$ eigenvalues are
 $\pm\lambda_1(X_N),\ldots,\pm\lambda_m(X_N)$.
 Furthermore $\lambda_1(X_N)^2,\ldots,\lambda_m^2(X_N)$ are the $m$ eigenvalues
 of $\frac{1}{N} U_C U_C\trans$, denoted in \cite{FRZ} as $Z(\tilde
 A_{n,m})$.  Clearly
 \begin{eqnarray}\nonumber
 \dete(\frac{\sqrt{t}}{\sqrt{N}}I_N-\i
 X_N)=(\frac{\sqrt{t}}{\sqrt{N}})^{n-m}
 \prod_{i=1}^m (\frac{t}{N}+\max(\lambda_i(X_N)^2,\epsilon)^2)\ge\\
 \det(\frac{\sqrt{t}}{\sqrt{N}}I_N-\i
 X_N)=(\frac{\sqrt{t}}{\sqrt{N}})^{n-m}
 \prod_{i=1}^m (\frac{t}{N}+\lambda_i(X_N)^2).\label{ineqdete}
 \end{eqnarray}
 Hence for $\epsilon\le 1$
 \begin{eqnarray*}
 0\le \frac{1}{N}(\log\dete(\frac{\sqrt{t}}{\sqrt{N}}I_N-\i
 X_N)-\log\det(\frac{\sqrt{t}}{\sqrt{N}}I_N-\i
 X_N))=\\
 \frac{1}{N}\sum_{\lambda_i(X_N)^2\le \epsilon^2}
 \log\frac{\frac{t}{N}+\epsilon^2}
 {\frac{t}{N}+\lambda_i(X_N)^2}\le
 \frac{1}{N}\sum_{\lambda_i(X_N)^2\le \epsilon^2}
 \log\frac{\epsilon^2} {\lambda_i(X_N)^2}\le\\
 \frac{1}{N}\sum_{\lambda_i(X_N)^2\le \epsilon^2}
 \log\frac{1} {\lambda_i(X_N)^2}.
 \end{eqnarray*}
 \cite[(3.2)]{FRZ} is equivalent to
 $$\limsup_{n\to\infty} \sup_{m\le n, \omega\in\Omega_{m,n,[b^2,a^2]}}
 {\rE}\frac{1}{m+n}\sum_{\lambda_i(X_{m+n})^2\le \epsilon^2_{m+n}}
 \log\frac{1} {\lambda_i(X_{m+n})^2}=0.
 $$
 Hence
 \begin{equation}\label{GZRi3}
 \lim_{N\to\infty}\frac{1}{N} ({\rE}\log \textrm{det}_{\epsilon_N}
 (\frac{\sqrt{t}}{\sqrt{N}}I_N-\i X_N)
 - {\rE}\log \det
 (\frac{\sqrt{t}}{\sqrt{N}}I_N-\i X_N))=0.
 \end{equation}
 Combine (\ref{ineqdete}) with Jensen's inequality to deduce
 $${\rE}\log \det(\frac{\sqrt{t}}{\sqrt{N}}I_N-\i X_N)\le
 \log  {\rE}\det(\frac{\sqrt{t}}{\sqrt{N}}I_N-\i X_N) \le
 \log  {\rE}\dete(\frac{\sqrt{t}}{\sqrt{N}}I_N-\i X_N)
 $$
 Hence
 \begin{eqnarray*}
 \limsup_{N\to\infty}\frac{1}{N}(\log  {\rE}\textrm{det}_{\epsilon_N}
 (\frac{\sqrt{t}}{\sqrt{N}}I_N-\i X_N)
 - {\rE}\log \det
 (\frac{\sqrt{t}}{\sqrt{N}}I_N-\i X_N)) \ge\\
 \limsup_{N\to\infty}\frac{1}{N}(\log  {\rE}\textrm{det}_{\epsilon_N}
 (\frac{\sqrt{t}}{\sqrt{N}}I_N-\i X_N) - \log {\rE}\det
 (\frac{\sqrt{t}}{\sqrt{N}}I_N-\i X_N))\ge 0.
 \end{eqnarray*}
 Use (\ref{GZRi0}) and (\ref{GZRi3}) to deduce (\ref{GZRi2}).  \qed

\end{document}